\documentclass[paper]{IEEEtran}
\usepackage{comment}
\usepackage{graphicx}
\usepackage{amsmath}
\usepackage{algorithm}
\usepackage{algorithmic}
\usepackage{subcaption}
\usepackage[numbers]{natbib} 
\usepackage{hyperref}
\usepackage{flushend}
\usepackage{authblk}
\usepackage{tablefootnote}
\usepackage{threeparttable}
\usepackage{stfloats}
\usepackage{svg}
\usepackage{comment}
\usepackage{pifont}
\title{ConSmax: Hardware-Friendly Alternative Softmax with Learnable Parameters}
\begin{document}

\author{\IEEEauthorblockN{Shiwei Liu$^{1,2 \dagger}$,
Guanchen Tao$^2$,
Yifei Zou$^2$, 
Derek Chow$^1$, 
Zichen Fan$^2$,
Kauna Lei$^2$,
Bangfei Pan$^1$,
Dennis Sylvester$^2$,
Gregory Kielian$^{1\ddagger}$, and
Mehdi Saligane$^{2\ddagger}$} \\
\IEEEauthorblockA{$^1$Google Research} \\
\IEEEauthorblockA{$^2$Department of Electrical Engineering and Computer Sciences, University of Michigan}
\thanks{
$\dagger$Work done during an internship at Google Research.
$\ddagger$The corresponding authors are Gregory Kielian (gkielian@google.com) and Mehdi Saligane (mehdi@umich.edu).}
}

\maketitle
\begin{abstract}
The self-attention mechanism distinguishes transformer-based large language models (LLMs) apart from convolutional and recurrent neural networks.
Despite the performance improvement, achieving real-time LLM inference on silicon remains challenging due to the extensive use of Softmax in self-attention.
In addition to the non-linearity, the low arithmetic intensity significantly limits processing parallelism, especially when working with longer contexts.
To address this challenge, we propose Constant Softmax (ConSmax), a software-hardware co-design that serves as an efficient alternative to Softmax.
ConSmax utilizes differentiable normalization parameters to eliminate the need for maximum searching and denominator summation in Softmax.
This approach enables extensive parallelization while still executing the essential functions of Softmax.
Moreover, a scalable ConSmax hardware design with a bitwidth-split look-up table (LUT) can achieve lossless non-linear operations and support mixed-precision computing.
Experimental results show that ConSmax achieves a minuscule power consumption of 0.2mW and an area of 0.0008mm$\rm ^2$ at 1250MHz working frequency in 16nm FinFET technology. 
For open-source contribution, we further implement our design with the OpenROAD toolchain under SkyWater’s 130nm CMOS technology.
The corresponding power is 2.69mW and the area is 0.007mm$\rm ^2$.
ConSmax achieves 3.35$\times$ power savings and 2.75$\times$ area savings in 16nm technology, and 3.15$\times$ power savings and 4.14$\times$ area savings with the open-source EDA toolchain.
In the meantime, it also maintains comparable accuracy on the GPT-2 model and the WikiText103 dataset.
The project is available at https://github.com/ReaLLMASIC/ConSmax.

\end{abstract}
\section{Introduction}

Transformer-based LLM models have become foundational across a wide range of machine learning domains, including natural language processing \cite{gpt, lang1} and computer vision \cite{cv0}.
The notable improvement can be attributed to the unique self-attention mechanism.
Different from previous convolutional or recurrent algorithms (CNN and RNN),
Self-attention mechanisms enhance LLMs' ability to capture information across input contexts (i.e., tokens) regardless of their distance.
Accelerating LLM inference on silicon is challenging due to its low arithmetic intensity, which results in poor energy efficiency. It prevents the further applications of LLMs, particularly on edge devices.

\begin{figure}[t]
\centering
\includegraphics[width=0.42\textwidth]{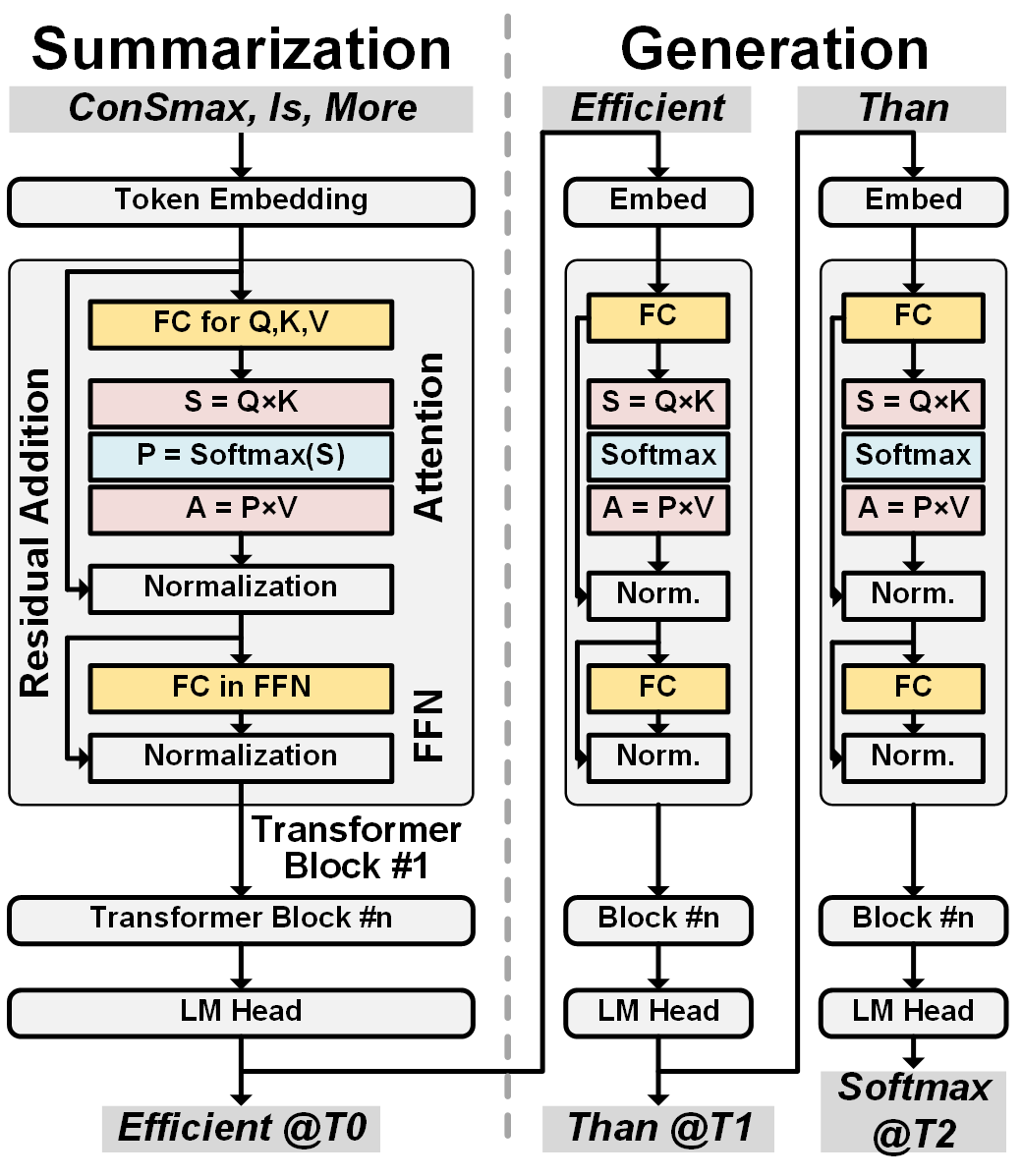}
\caption{LLM structure and processing flow with summarization and generation stages.\label{fig:llm}}
\vspace{-4mm}
\end{figure}

Softmax is the primary factor exacerbating the inefficiency of LLMs when processing long contexts.
Fig.~\ref{fig:llm} provides a detailed illustration of the LLM structure, including the self-attention mechanism within the model.
The self-attention first transforms the input token embeddings into separate representations known as Query (\textbf{Q}), Key (\textbf{K}), and Value (\textbf{V}).
Self-attention then consists of two matrix multiplications, with a Softmax normalization in between them.
The Q$\times$K multiplication generates attention scores (\textbf{S}), which indicates the relevance between different tokens.
The self-attention always extracts contextual information from highly-related tokens.
Subsequently, the Softmax function normalizes the score matrix to derive attention probabilities (\textbf{P}).
Finally, the P$\times$V multiplication calculates a weighted average of all value vectors, producing the self-attention output.
Previous works \cite{a3, elsa, spatten} primarily focus on optimizing the matrix multiplications in self-attention, but often overlook the Softmax operation.
However, Softmax can be the actual bottleneck in LLM inference.
Computing Softmax requires allocating and iterating over the entire score vector to determine the maximum score and the score summation.
Consequently, the after-Softmax P$\times$V should be blocked until the before-Softmax Q$\times$K and Softmax are completed.
Due to the low parallelism opportunity, Softmax reduces the hardware utilization of mainstream GPUs and TPUs, as well as other potential hardware.
The situation exacerbates when self-attention processes longer contexts, as the Softmax complexity increases linearly with the token numbers.
Softermax \cite{softermax} reveals that the Softmax can lead to over 30\% latency overhead during LLM inference,
particularly with a token sequence exceeding 4K, a typical size in state-of-the-art LLMs \cite{gpt, mistral, llama}.

While many previous designs exploit quantization \cite{pquant, gptq, quip}, weight sparsity \cite{sparsegpt, icarus}, token relevance sparsity \cite{a3, elsa, spatten, sprint, fact, dota} and workload partition \cite{transpim, dfx, flashattention} to optimize the matrix multiplications in LLM,
these endeavours are still hindered by the Softmax bottleneck.
A few pioneers \cite{softermax, flashdecoding++, base2} focus on Softmax optimization.
They can be divided into two categories, namely computing approximation and workflow scheduling.
For computing approximation, Look-up tables (LUTs) \cite{lut0, nulut, lut2} and Taylor expansions \cite{taylor0, taylor1} are two methods to approximate the non-linear Softmax. 
However, approximate computing invariably compromises LLM accuracy and fails to enhance Softmax parallelism.
Conversely, workload scheduling prioritizes enhancing Softmax parallelism.
For example, SpAtten \cite{spatten} computes Q$\times$K, Softmax and P$\times$V for a sequence of tokens in pipeline.
While SpAtten demonstrates high efficiency on encoder-only BERT models \cite{bert},
it still suffers from a low utilization on decoder-only GPT models,
which are the representative of the prevailing LLMs.
On the other hand, partial Softmax, such as FlashAttention \cite{flashattention, flashattention2}, increases Softmax parallelism for both encoder and decoder LLMs.
The central concept involves partitioning the score vector into multiple partial vectors and subsequently applying standard Softmax on each of them in parallel.
Nevertheless, partial Softmax requires synchronization across partial vectors to determine the ultimate maximum score and score summation,
which accounts for about 20\% runtime latency for self-attention computing \cite{flashdecoding++}.

In summary, previous Softmax-oriented works suffer from inefficiency stemming from a lack of high parallelism and accuracy.
There is a noteworthy observation that the maximum score can be replaced by arbitrary values to scale the numerator and denominator in Softmax.
Furthermore, the probability vector resulting from Softmax normalization does not necessarily have to be a unit vector to maintain LLM accuracy.
Inspired by the above insights, this paper proposes Constant Softmax (ConSmax), a software-hardware co-design that serves as an efficient alternative to Softmax.
Diverging from previous strategies,
ConSmax enhances computing parallelism and ensures lossless non-linear computing.
The key contributions are listed as follows:
\begin{itemize}
\item We utilize two differentiable normalization parameters to substitute for the maximum score and denominator in the original Softmax, 
thereby preventing the data synchronization required for maximum searching and score summation.
These parameters are learnable during training and remain fixed during inference, thus achieving inference efficiency.

\item We propose a bitwidth-split ConSmax hardware design to generate lossless non-linear functions and mitigate the lookup table (LUT) overhead.
Furthermore, the ConSmax hardware is scalable to accommodate mixed-precision computing, a prevalent feature in state-of-the-art LLMs \cite{mistral, mix}.

\item We extensively evaluate ConSmax on the GPT-2 model and the WikiText103 \cite{wikitext103} dataset.
Experimental results show that ConSmax achieves 3.35$\times$ power savings and 2.75$\times$ area savings in 16nm FinFET technology, and 3.15$\times$ power savings and 4.14$\times$ area savings with the open-source OpenROAD toolchain under SkyWater’s 130nm CMOS technology.
In the meantime, ConSmax also maintains comparable accuracy on the GPT-2 model and the WikiText103 dataset.

\end{itemize}

\section{Background}
\label{sec:background}

\subsection{Large Language Model Preliminaries}
\label{sec:llm}

\subsubsection{\textbf{Structure}}
Fig~\ref{fig:llm} illustrates the LLM structure.
It typically contains three main architectural components: the embedding layer, the transformer block and the language model (LM) head.
The embedding layer consists of token embedding and positional encoding, encoding the discrete input tokens to high-dimensional representations.
Token embedding captures the semantic meaning, while positional encoding records the relative positioning order.
In contrast, at the very end is the LM head, which serves the converse function to the embedding layer.
It takes in the transformer outputs and transforms them back into linguistic tokens by predicting the probabilities of the next token.
Connecting the embedding layer and LM head is a stack of transformer blocks, constituting the bulk of the LLMs.
Each block can be further divided into a multi-head self-attention layer and a feed-forward layer.
The self-attention layer allows each token to attend to every other token,
thus enabling the model to capture global information across input tokens regardless of their distance.
The self-attention is applied multiple times to form multi-head self-attention.
All attention heads operate in parallel to capture linguistic dependencies from various representation subspaces.
The feed-forward layer further processes the attention output through two linear transformations, 
providing additional representational capacity to LLM models.

\subsubsection{\textbf{Workflow}}

As shown in Fig.~\ref{fig:llm}, for text generation tasks, LLMs operate between a summarization stage and a generation stage.
The summarization stage provides the initial prompt context to condition the LLM model, while the generation stage uses the context to produce a continuation.
Both stages utilize the same model structure but with distinct workflows.
In the summarization stage, the self-attention layer simultaneously processes a set of input tokens and extracts their key and value representations in parallel. 
The resultant representation matrix is then reused in the generation stage to generate the first new token.
In contrast, the generation stage generates only a single token at each inference iteration.
Subsequently, the output tokens from the previous iteration are iteratively fed back as input to generate subsequent output tokens.

\subsection{Softmax Bottleneck in LLM Inference}
\label{sec:softmax}

As the summarization stage processes a set of input tokens, it involves matrix-matrix multiplications, rendering it compute-bound.
Consequently, the summarization stage can effectively saturate GPUs/TPUs utilization.
In contrast, the generation stage performs vector-matrix multiplication to generate a single new token at each inference, which is a memory-bound operation. 
This can result in a reduced utilization of GPUs/TPUs or even previous transformer-oriented accelerators \cite{{elsa, spatten, sprint, fact, dota}}.
Softmax leads to the underutilization in the generation stage.
Unfortunately, this aspect has not been fully explored in these prior works.

\begin{figure}[t]
\centering
\includegraphics[width=0.48\textwidth]{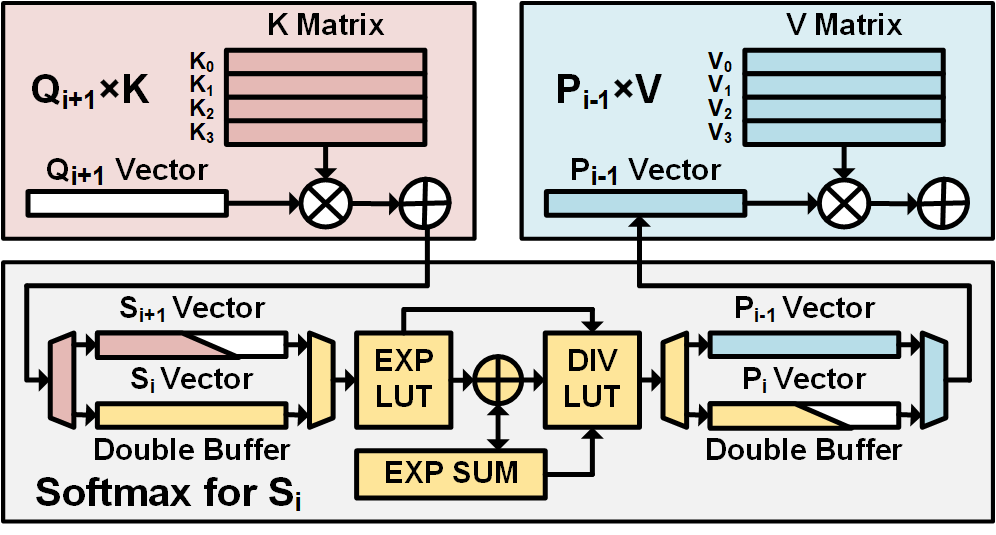}
\caption{Self-attention inference on hardware with token-pipeline workflow.\label{fig:token_pipeline}}
\vspace{-5mm}
\end{figure}

Fig.~\ref{fig:token_pipeline} illustrates the architecture of the typical transformer accelerators \cite{elsa, spatten}.
For brevity, the peripheral modules such as the sparsity detector in these works are not depicted.
These types of accelerators process Q$\times$K, Softmax and P$\times$V in a pipeline manner.
Note that the original Softmax requires data synchronization to calculate the maximum factor and denominator.
Therefore, for each token, the Softmax module should allocate all Q$\times$K results and halt the P$\times$V computation until Softmax is completed.
To enhance computing parallelism, these cascaded modules in Fig.~\ref{fig:token_pipeline} process different tokens in a pipeline, known as the token-pipeline workflow.
For example, Q$\times$K module operates on Token$\rm _{i+1}$, while Softmax and P$\times$V modules handle Token$\rm _{i}$ and Token$\rm _{i-1}$, respectively.
Within each operation, the partial result is stored in double buffering and interleaved to the next module once the final results are ready.
Token-pipeline flow facilitates the summarization stage, during which LLM models process plenty of tokens from the input prompt.
However, as mentioned in Section~\ref{sec:llm}, the generation stage cannot fully utilize the pipeline because the LLMs operate with only a single input token.
As a result, only one of the three modules works at a time, causing a decrease in hardware utilization.

In addition to its low parallelism, Softmax also faces challenges due to hardware-unfriendly non-linear operations such as exponents and reciprocals.
To implement Softmax on silicon, previous works utilize LUTs as well as Taylor expansions to approximate Softmax as piece-wise linear functions.
These methods are effective for accelerating CNNs and RNNs, where Softmax accounts for a minimal portion as the final classification layer.
However, this prerequisite is no longer valid for transformer-based LLM models, which employ Softmax as the key component in the self-attention mechanism.
On the one hand, using approximated Softmax can compromise LLM accuracy.
On the other hand, it does not enhance computing parallelism.

\subsection{ConSmax Motivation}

In summary, the Softmax operation hinders the existing LLM accelerators due to its low computing parallelism and hardware-unfriendly non-linear operations.
To address these challenges, we propose ConSmax, a software-hardware co-design that serves as an efficient alternative to Softmax.
Our work is orthogonal to previous works, contributing:

\subsubsection{\textbf{High Computing Parallelism}} 
Rather than focusing on the already optimized matrix multiplication,
we thoroughly investigate the Softmax bottleneck in LLM acceleration.
The maximum searching and denominator summation contribute to approximately 20\% of the latency in the attention operation during token generation \cite{flashdecoding++}.
To mitigate this synchronization overhead,
ConSmax introduces differentiable normalization parameters that replace the maximum factor and denominator in the original Softmax,
enhancing computing parallelism by eliminating the need for maximum score searching and score summation.

\subsubsection{\textbf{Lossless Computing and Scalability}}
A bitwidth-split ConSmax hardware can perform lossless non-linear operations while minimizing LUT overhead.
Additionally, the ConSmax hardware is hierarchical and scalable, supporting mixed-precision computing, which is commonly used in state-of-the-art LLM models \cite{mistral}.
\section{ConSmax Algorithm}
\label{sec:algorithm}

\begin{figure*}[ht]
\centering
\includegraphics[width=0.75\textwidth]{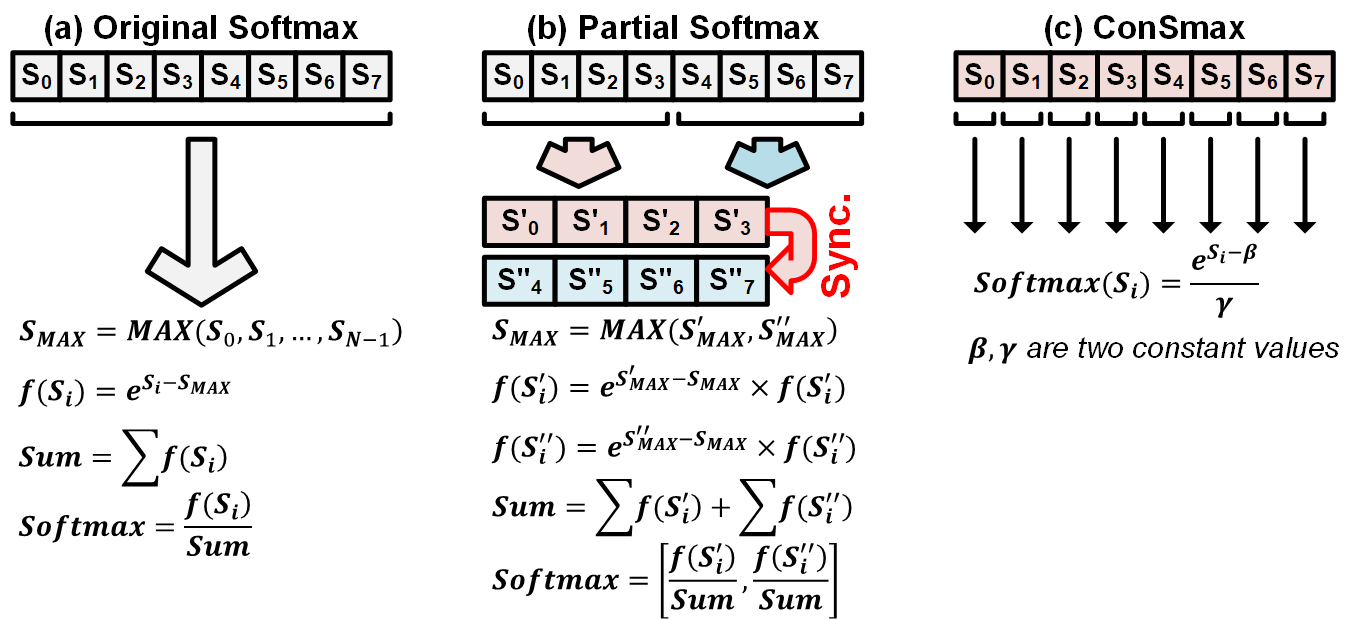}
\caption{Comparing ConSmax with (a) original Softmax and (b) partial Softmax. \label{fig:consmax_software}}
\vspace{-3mm}
\end{figure*}

Although the Softmax operation represents a relatively small portion of the overall self-attention layer,
it poses a much greater challenge to be designed into efficient hardware compared to well-optimized matrix multiplication.
Therefore, the Softmax can result in significant overhead if not handled appropriately.
In this section, we introduce the ConSmax algorithm, specifically designed to alleviate data synchronization requirements for computing the exponential maximum and summation within the original Softmax.
As a consequence, ConSmax significantly enhances computing parallelism during LLM inference, notably benefiting the generation stage.

\subsection{Convert Softmax to ConSmax}

The Softmax operation depicted in Fig.~\ref{fig:consmax_software}(a) requires the entire score vector to determine the maximum score and score summation.
However, generating all scores simultaneously in a single cycle is impractical,
particularly considering the current LLM models that accommodate thousands to hundreds of thousands of input tokens. \cite{gpt, mistral, llama} (from 8K to 128K tokens).
Therefore, the potential hardware should buffer the exponent results for all score elements before Softmax can proceed.
Note that the exponent operation would explode the numerical range. 
It results in more memory consumption to maintain the precision of these intermediate results.
To increase Softmax parallelism, \emph{\textbf{the primary challenge lies in computing each partial Softmax result independently, without relying on the maximum score and summation results from other partial Softmax computations.}}

According to the mathematical formula, Softmax operation works as a normalization function, 
wherein the maximum value serves as the scaling factor to prevent data overflow.
There is a noteworthy observation that the scaling factor can be an arbitrary value rather than strictly using the maximum value. 
Therefore, the Softmax operation is converted as:

\begin{equation}
Softmax(S_i) = \frac{e^{S_i-S_{max}}}{\sum_{i}{e^{S_i-S_{max}}}}= \frac{e^{S_i-\beta}}{\sum_{i}{e^{S_i-\beta}}}
\end{equation}

where S$\rm _i$ represents the i-th score element and $\beta$ could be arbitrary value.
Further, the denominator could also be arbitrary as long as it satisfies the normalization function.
Therefore, we replace the denominator with another constant value, denoted as $\gamma$, to propose the final ConSmax formulation as:

\begin{equation}
ConSmax(S_i) = \frac{e^{S_i-\beta}}{\gamma}
\label{eq:consmax}
\end{equation}

However, it's essential to note that the scaling factor $\beta$ cannot be arbitrary due to the risk of exponential computation overflow.
For example, the overflow occurs in the case where S$\rm _i >> \beta$.
In contrast, the exponential result infinitely tends to be zero if S$\rm _i << \beta$, leading to numerical precision loss.
Similarly, the denominator $\gamma$ cannot be chosen arbitrarily, since it normalizes the exponential score to probability distribution.
Take the most extreme case as an example, where $\gamma\rightarrow0$ or $\gamma\rightarrow+\infty$. 
In such cases, the probability values tend towards to infinity or zero, respectively, rendering them unable to effectively distinguish the token relevance.

To determine the optimal scaling and denominator, we designate the $\beta$ and $\gamma$ as learnable parameters.
During the training phase, these parameters evolve in response to the characteristics of the practical dataset.
In addition, the combination of $\beta$ and $\gamma$ varies across different self-attention heads,
allowing for a more flexible and customized approach to normalization.
The next problem is how to initialize the $\beta$ and $\gamma$ before training commences.
Optimizing $\beta$ and $\gamma$ is pivotal for enhancing the effectiveness of the ConSmax function, 
presenting a promising avenue for improving LLM efficiency.
This exploration can be conducted through a hyperparameter tuning process,
where various combinations of initial $\beta$ and $\gamma$ are tried during the warming-up iterations.
The combination that yields the best performance, such as the lowest validation loss, is selected.

Finally, while a pretrained denominator cannot guarantee that the probability vector is a unit vector,
the experimental results presented in Section~\ref{sec:result} demonstrate that this constraint does not degrade LLM accuracy.
As long as the probability distribution can magnify the small differences in input scores, the LLM performance remains robust.
In addition, the ConSmax in Equation~\ref{eq:consmax} can be rewritten as:

\begin{equation}
ConSmax(S_i) = \frac{e^{S_i-\beta}}{\gamma} = C \times e^{S_i}, where \hspace{1mm} C = -\frac{e^{\beta}}{\gamma}
\label{eq:inference}
\end{equation}

During inference, we merge $\beta$ and $\gamma$ into a single constant value.
However, during training, we maintain them as independent parameters to mitigate against exponential overflow.

\subsection{Comparison with Partial Softmax}

Softermax \cite{softermax} and FlashDecoding++ \cite{flashdecoding++} are two alternative approaches to Softmax, aimed at reducing memory consumption and enhancing computational parallelism.
Similarly, they employ the partial Softmax technique, where the global maximum factor is replaced by the local one.
More specifically, the main idea is to divide the score vector into several partial vectors and subsequently apply the standard Softmax to each partial vector separately.
Although these approaches improve Softmax parallelism, they necessitate additional synchronization among different partial vectors.
As shown in Fig.~\ref{fig:consmax_software}(b), the partial Softmax recalculates the exponential maximum and summation after all partial vectors are processed by the standard Softmax. Subsequently, their normalized results are adjusted based on the synchronized maximum and summation values.
Such synchronization leads to 18.8\% overheads in the attention computation with 1024 input tokens, which is anticipated to worsen with longer input sequences.
In contrast, ConSmax avoids any synchronization, distinguishing itself from these previous works.
\section{ConSmax Hardware}
\label{sec:hardware}

\begin{figure*}[t]
\centering
\includegraphics[width=0.85\textwidth]{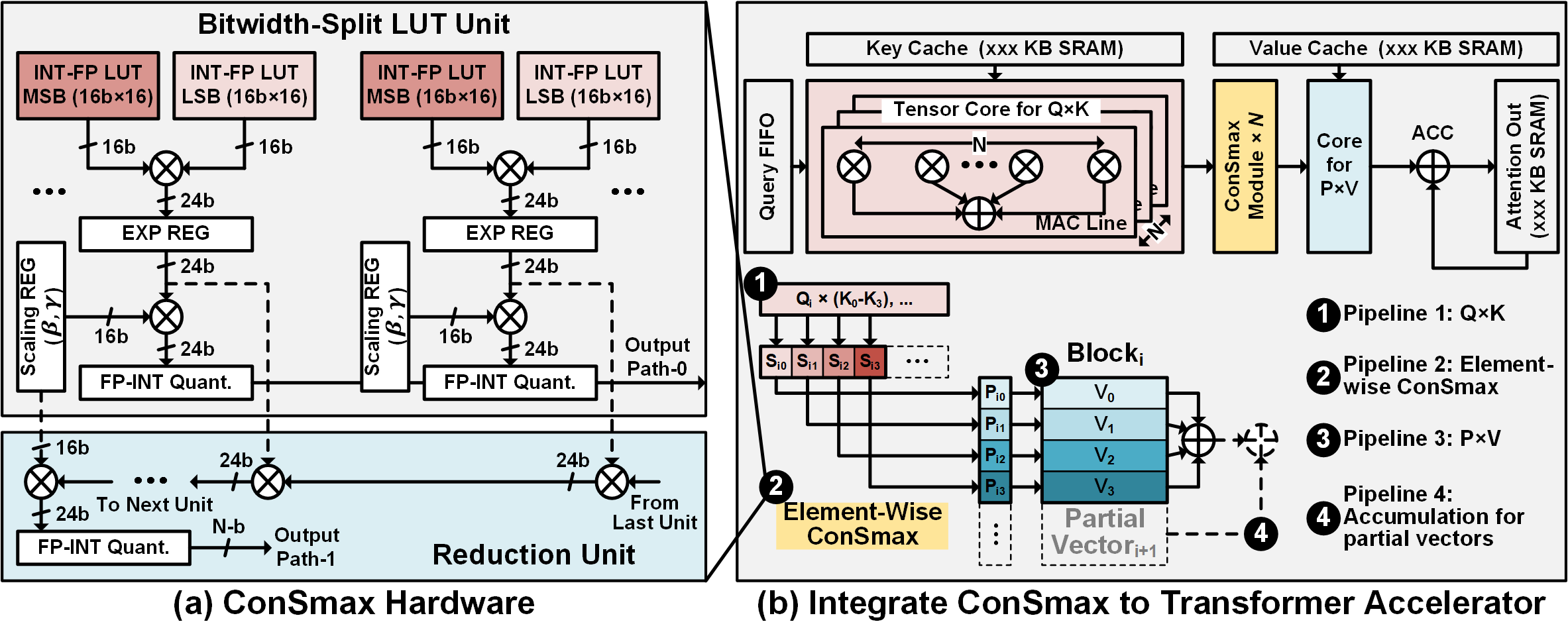}
\caption{(a) Bitwidth-split ConSmax hardware unit and (b) Integrate ConSmax hardware to transformer accelerator. \label{fig:consmax_hardware}}
\vspace{-4mm}
\end{figure*}

Based on the algorithm in Section~\ref{sec:algorithm}, 
we proceed to explore the customized ConSmax hardware.
The bitwidth-split LUT architecture can produce lossless non-linear functions and enable the scalability for the mix-precision computing.
In addition, we also elaborate on how to integrate the ConSmax hardware into state-of-the-art transformer accelerators.
Leveraging its synchronization-free property, ConSmax enhances parallelism in both LLM summarization and generation stages.

\subsection{Lossless and Scalable ConSmax Hardware}

Fig.~\ref{fig:consmax_hardware}(a) illustrates the proposed ConSmax hardware featuring a two-level structure.
In Level-1, multiple bitwidth-split ConSmax units operate in parallel to execute ConSmax operations.
The reduction unit in Level-2 can allocate varying numbers of the basic ConSmax units to support mixed-precision computing.
For brevity, Fig.~\ref{fig:consmax_hardware}(a) just displays two ConSmax units, capable of generating either two 8-bit ConSmax results or one 16-bit result simultaneously.

\subsubsection{\textbf{Bitwidth-Split ConSmax Unit}} As depicted in Fig.~\ref{fig:consmax_hardware}(a), each ConSmax unit comprises four components as bitwidth-split LUTs, floating-point (FP) multipliers, FP-to-Integer (INT) converter and necessary buffers.
Assume that matrix multiplication accelerators, such as GPUs/TPUs, produce the Q$\times$K multiplication and generate 8b-INT scores sequentially.
Upon receiving the 8b-INT results, the LUTs within the ConSmax unit first perform the exponential operation and concurrently dequantize the result to 16b-FP format.
Specifically, the 8b-INT scores function as addresses to access the LUTs, which store the corresponding 16b-FP exponential/dequantized results.
This approach allows for storing the exponential result within a limited bitwidth while maintaining higher precision,
thus mitigating potential bitwidth explosion inherent in integer format representation.
Moreover, the ConSmax unit circumvents the need for an additional INT-to-FP converter, thanks to the simultaneous conversion by the LUTs.
To further minimize the LUT capacity, instead of employing a large LUT to enumerate all 256 combinations, we split the 8-bit input into two slices as MSB-INT4 and LSB-INT4.
Each fragment is associated with a 16-entry LUT.
The partial sums from these two LUTs are then merged in the downstream multiplier.
This configuration allows the bitwidth-split LUTs to perform lossless exponential operations for all input combinations with minimal LUT overhead.
In contrast to the straightforward shift and addition used for integral bitwidth alignment, the partial sum reduction between floating-point fragments is comparatively intricate:

\begin{equation}
\label{eq:reduction}
\begin{aligned}
    e^{S_{INT8}} &= e^{(MSB_{INT4} << 4) + LSB_{INT4}} \\ 
                 &= e^{2^{4} \times MSB_{INT4}}\times e^{LSB_{INT4}}
\end{aligned}
\end{equation}

To avoid the implementation of non-linear $(e)^{2^4}$ in hardware,
the MSB-LUT directly projects $e^{2^{4}\cdot x}$ for MSB-INT4, while LSB-LUT exclusively maps $e^x$.
Following exponent calculation, the ConSmax normalization is further applied according to Equation~\ref{eq:consmax}.
The pretrained $\beta$ and $\gamma$ are combined based on Equation~\ref{eq:inference}and then multiplied with the exponential result in the second multiplier.
Based on the above analysis, compared to previous works \cite{lut0, nulut}, ConSmax can generate accurate nonlinear operations without using a LUT to approximate Softmax.

\subsubsection{\textbf{Reduction Unit}} 

The mix-precision computing is a prevalent model compression technique for efficient LLM inference \cite{mistral},
wherein different operators of the model can be assigned with different precision.
Therefore, this feature necessitates the need for supporting mixed-precision computing in the ConSmax hardware.
To achieve this, the reduction unit allocates multiple bitwidth-split LUTs according to the precision requirements.
As shown in Figure~\ref{fig:consmax_hardware}(a),
two 8-bit bitwidth-split LUTs can execute one 16-bit ConSmax normalization.
The 16-bit score element is initially divided into two 8-bit slices.
Subsequently, each LUT receives one slice and independently generates an 8-bit ConSmax.
The partial sums from different LUTs are then directed to the reduction unit, where they are merged using a floating-point multiplier chain, as depicted in Equation~\ref{eq:reduction}.
The reduction unit modifies the length of the multiplier chain to accommodate other precision configurations.

\subsection{Integrate ConSmax Hardware to Transformer Accelerator}

Figure~\ref{fig:consmax_hardware}(b) elaborates on the integration of ConSmax hardware into dedicated transformer accelerators.
Two TPU tensor cores (or alternative hardware, e.g. GPUs), alongside the inserted ConSmax hardware, produce Q$\times$K, ConSmax normalization and P$\times$V in pipeline.
Note that the most advanced LLMs typically support input contexts with $\geq$8K tokens (e.g. 32K in GPT-4\cite{gpt}).
Therefore, it is improbable for the tensor core to execute such extensive matrix multiplication simultaneously.
Nevertheless, thanks to ConSmax, the pipeline can continue to function effectively.
\ding{182} First, the front-end tensor core conducts Q$\times$K operation between the single given query vector and a portion of the key vectors.
The size of the involved key vectors scales proportionally to the capacity of the tensor core.
\ding{183} Secondly, the partial score vector is directly forwarded to the ConSmax unit for normalization.
Meanwhile, the front-end tensor core simultaneously computes the Q$\times$K multiplications for the next set of key vectors.
\ding{184} Despite not having generated the complete attention probabilities,
the normalized elements can be directly multiplied with the corresponding value vectors in the back-end tensor core.
This is facilitated by ConSmax, which eliminates the need for the exponential maximum and summation calculations in the original Softmax function.
\ding{185} Finally, the partial sum is accumulated and updated to generate the attention output.

\begin{figure}[t]
\centering
\includegraphics[width=0.48\textwidth]{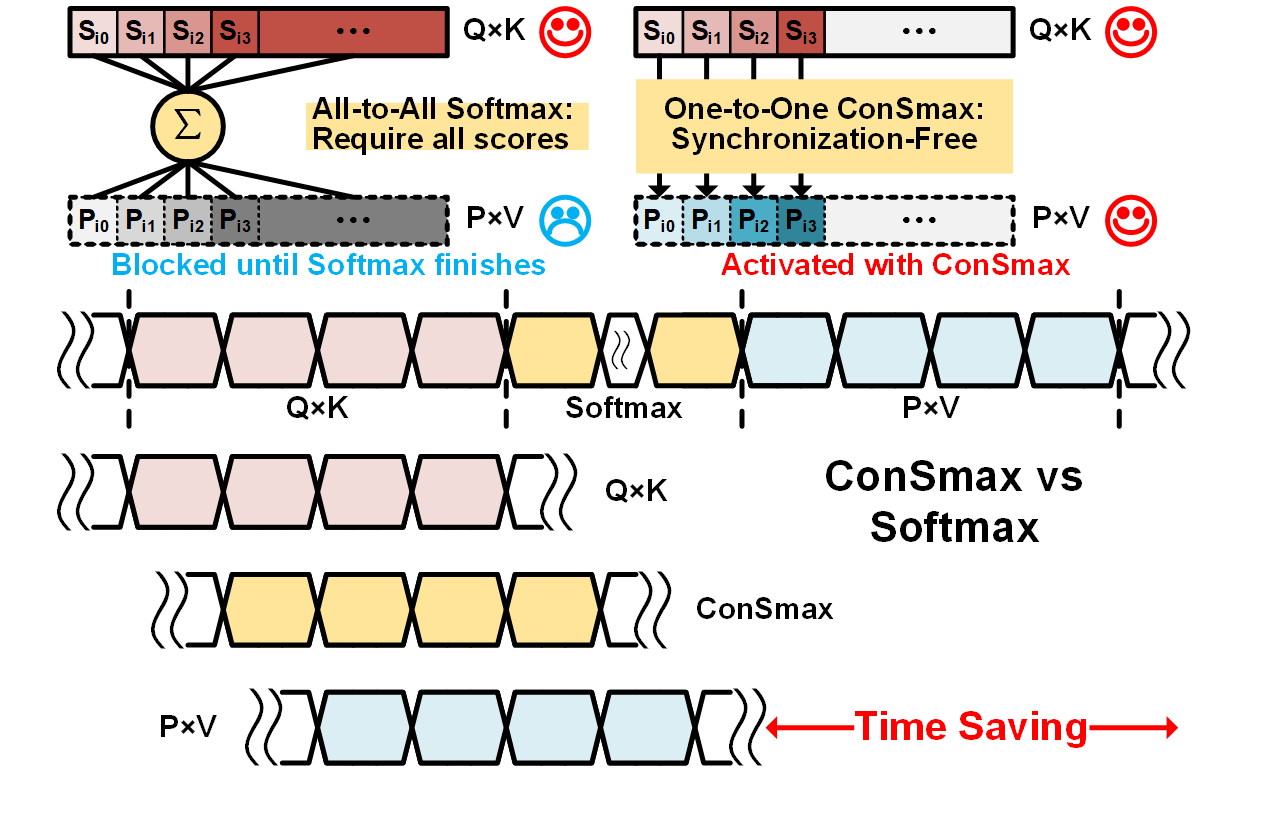}
\vspace{-2mm}
\caption{Synchronization-free ConSmax leads to high parallelism and time savings. \label{fig:time_saving}}
\vspace{-6mm}
\end{figure}

In contrast to the coarse-grained token pipeline outlined in Section~\ref{sec:softmax},
ConSmax-integrated accelerators can employ a fine-grained element-wise pipeline to attain higher parallelism and realize significant time savings.
Note that the LLM generation stage comprises only a single input token, which is insufficient to fully utilize the pipelined Q$\times$K and P$\times$V hardware modules in Figure~\ref{fig:token_pipeline}.
Therefore, as shown in Figure~\ref{fig:time_saving}, the P$\times$V operation in such accelerators is stalled until the original Softmax is completed, thereby leading to significant underutilization.
The accelerator can allocate all computing logic to process LLMs in a layer-sequential manner \cite{ls},
thereby achieving high utilization, 
However, this approach could result in numerous intermediate results which necessitate a large storage overhead.
On the contrary, with its synchronization-free attribute, ConSmax allows for an element-wise pipeline instead of a token-based one.
Consequently, the accelerator can fully utilize all processing modules even with a single token.
This not only accelerates LLM summarization but, more significantly, enhances LLM generation with a uniform architecture.
\section{Experimental Results}
\label{sec:result}

\subsection{Experiment Setup}

We have developed a ConSmax prototype using Verilog RTL under 16nm FinFET CMOS technology.
All digital modules are synthesized using Synopsys Design Compiler.
For open-source contribution, we further synthesized our design with the OpenROAD toolchain \cite{openroad} under SkyWater's 130nm CMOS technology.
These resulting netlists are used to evaluate the power and area consumption.
We compare the ConSmax hardware with two baselines: the  DesignWare-based Softmax hardware and Softermax \cite{softermax}.
The former Softmax hardware faithfully implements the original Softmax function, 
whereas the Softermax is recognized as an efficient alternative to the conventional Softmax function.
Since Softermax does not explicitly provide power and area results,
we develop a corresponding counterpart according to the Softermax algorithm.
These baselines are synthesized with the same configurations.

We evaluate the impact of ConSmax on the GPT model using the WikiText103 \cite{wikitext103} dataset, where the Softmax in each self-attention block is replaced by the ConSmax.
This benchmark model contains 6 transformer layers each equipped with 6 self-attention heads.
Therefore, the embedding size is set to 384.
Moreover, the default token length is set to 256.
We report perplexity for the WikiText103 dataset.
This metric measures the LLM's performance on text generation tasks.
A lower perplexity value indicates better performance.

\subsection{Software Performance}

\begin{figure}[t]
\centering
\includegraphics[width=0.48\textwidth]{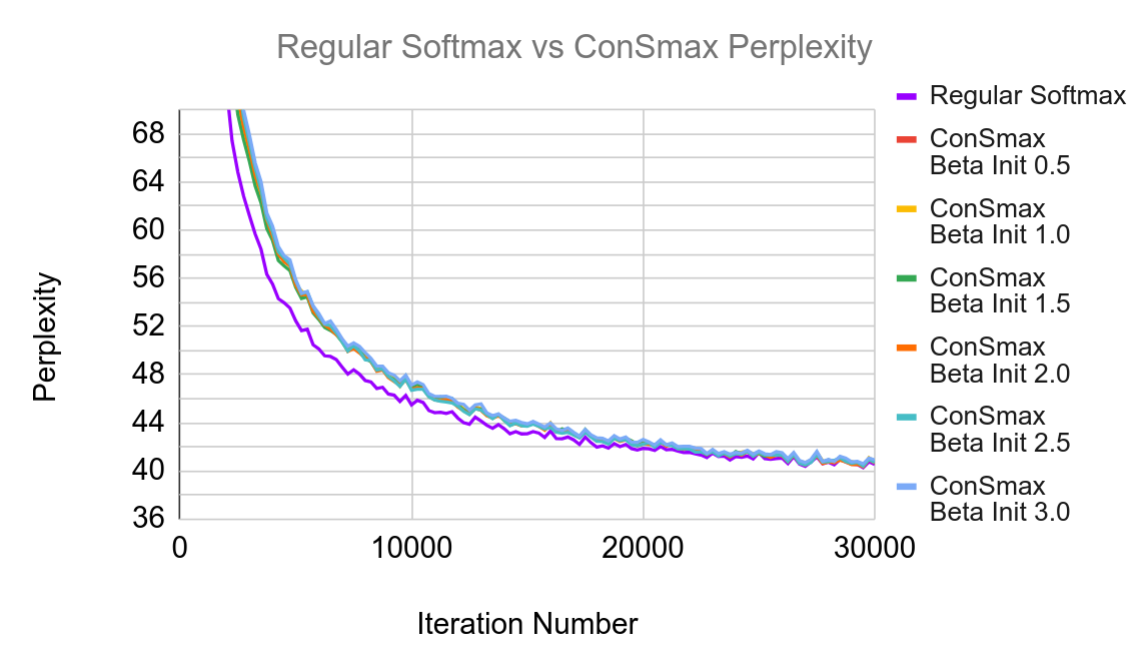}
\caption{Perplexity of GPT-2 models with Softmax and ConSmax, showing convergence of validation losses. \label{fig:loss}}
\vspace{-3mm}
\end{figure}

\begin{figure}[t]
\centering
\includegraphics[width=0.48\textwidth]{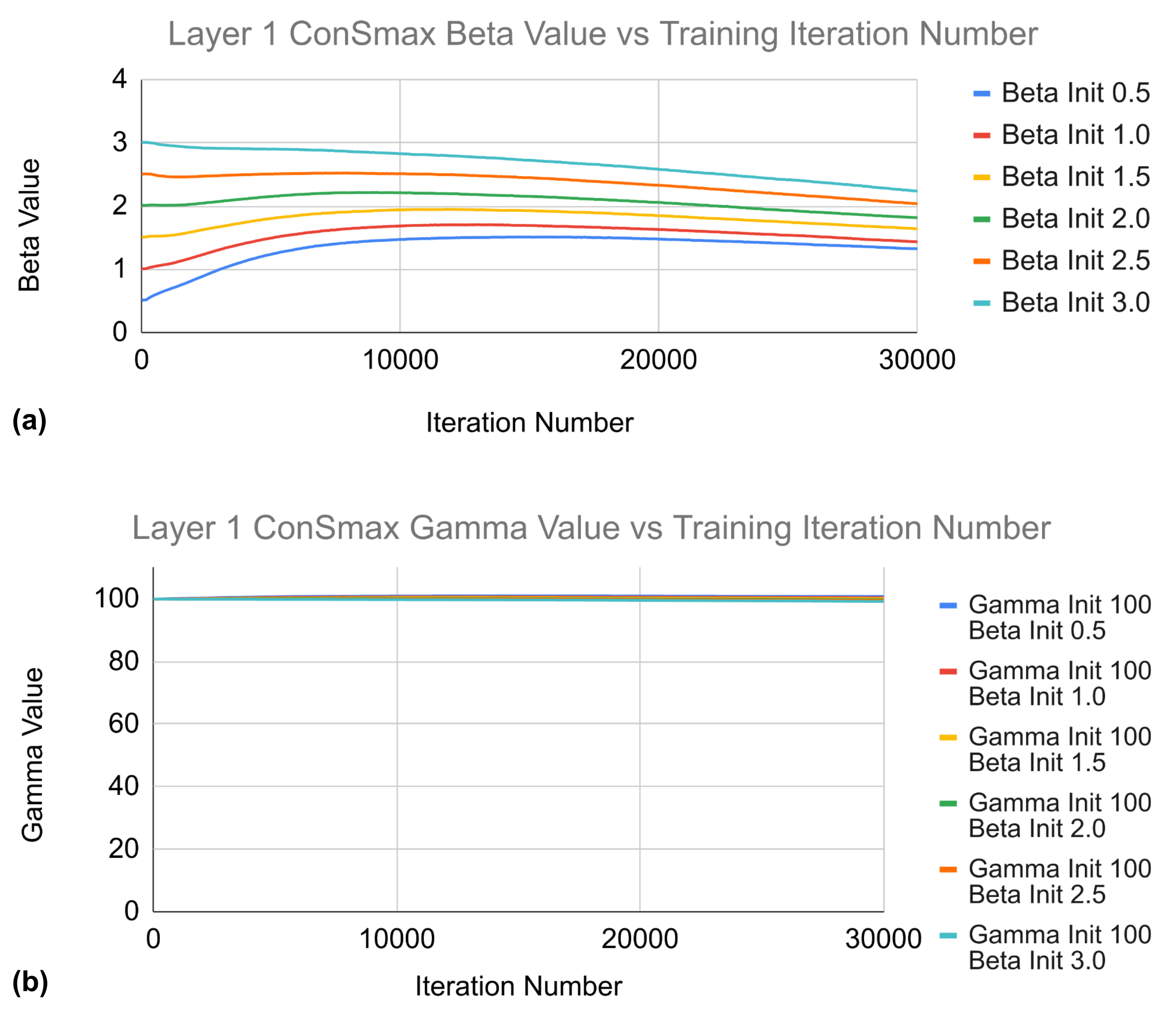}
\caption{Evolution of $\beta$ and $\gamma$ throughout training, each training run using a different starting value for $\beta$ (as $\gamma$ has been observed to have low \% change). Spread of $\beta$ values has been found to decrease with training. \label{fig:beta_gamma}}
\vspace{-3mm}
\end{figure}

\begin{figure}[t]
\centering
\includegraphics[width=0.44\textwidth]{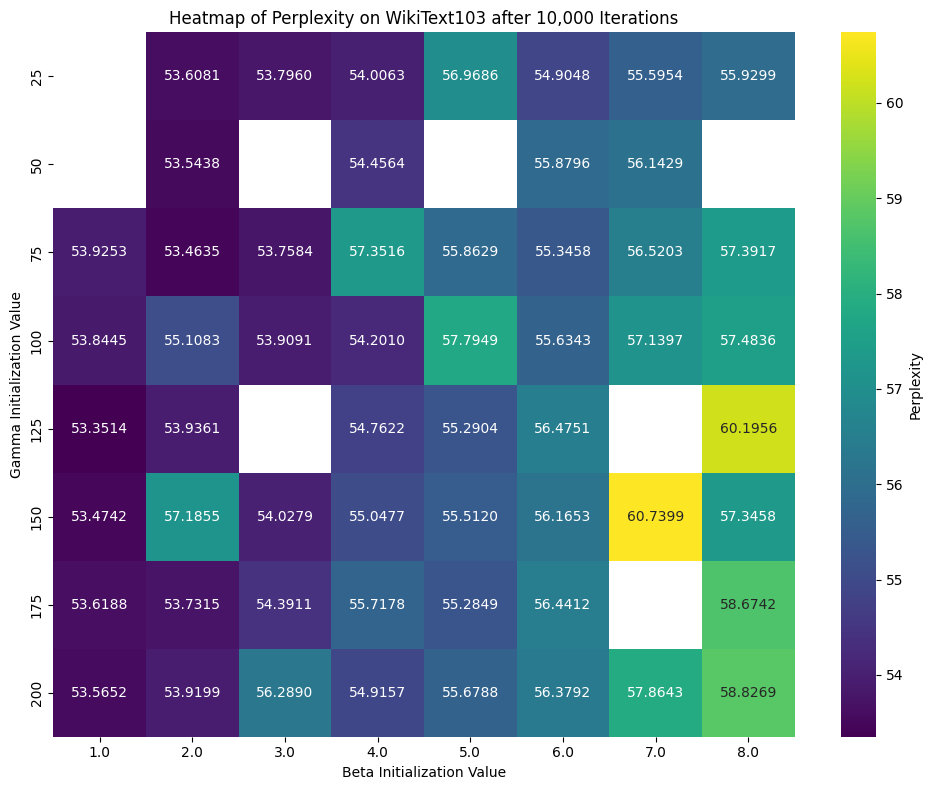}
\caption{Explore $\beta$ and $\gamma$ for ConSmax-based LLM training. \label{fig:explore}}
\vspace{-3mm}
\end{figure}

\begin{table*} [t]
  \centering
  \begin{threeparttable}
    \caption{ConSmax Hardware Performance Comparison with Softermax, Softmax}
    \begin{tabular}{|l|c|c|c|c|c|c|} 
    \hline
    \textbf{Proprietary EDA} & ConSmax & Softermax & Softmax & Consmax & Softermax & Softmax\\
    \hline
    Process  & \multicolumn{3}{c|}{16nm} & \multicolumn{3}{c|}{130nm} \\
    \hline
    Max Frequency (MHz) & 1250 & 1111 & 909 & 666.67 & 333.33 & 285.71 \\
    \hline
    Area (mm$^2$)\tnote{a} & 0.0008 & 0.0022 & 0.011 & 0.007 & 0.029 & 0.18 \\
    \hline
    Power (mW)\tnote{b} & 0.2 & 0.67 & 1.5 & 2.69 & 8.5 & 51 \\
    \hline
    Optimum Energy per op(pJ) & 0.2 & 0.7 & 1.5 & 4 & 25.5 & 178.5 \\
    \hline
  \end{tabular}
  
  \begin{tabular}{|l|c|c|c|c|c|c|} 
    \hline
    \textbf{Opensource EDA} & ConSmax & Softermax & Softmax & Consmax & Softermax & Softmax\\
    \hline
    Process  & \multicolumn{3}{c|}{16nm} & \multicolumn{3}{c|}{130nm} \\
    \hline
    Max Frequency (MHz) & 2000 & 1000 & 500 & 166.67 & 142.86 & 87.72 \\
    \hline
    Area (mm$^2$)\tnote{a} & 0.0009 & 0.0019 & 0.011 & 0.015 & 0.033 & 0.2 \\
    \hline
    Power (mW)\tnote{b} & 0.683 & 2.82 & 10.2 & 1.82 & 10.5 & 42.2 \\
    \hline
    Optimum Energy per op(pJ) & 0.3 & 1.4 & 2.7 & 16.7 & 73.5 & 255.3 \\
    \hline
  \end{tabular}
  
  \begin{tablenotes}
    \item[a] Power for 16nm is tested with 500MHz, for 130nm is tested with 80MHz.
    \item[b] Area is measured at Max Frequency.
  \end{tablenotes}
    \label{table:hardware_comparison}
  \end{threeparttable}
\vspace{-5mm}
\end{table*}

We first compare the perplexity of the standard Softmax and the proposed ConSmax.
In this experiment, we initialize $\beta$ within the range of [0.5, 2.5], while $\gamma$ is set to a constant value of 100.
As shown in Figure~\ref{fig:loss}, with varying combinations of $\beta$ and $\gamma$, 
ConSmax initially exhibits a marginally 2.3\% higher perplexity than Softmax, and leads to less than 0.9\% perplexity degeneration after 10K iterations.
After about 20K iterations, both Softmax and ConSmax-based models converge, demonstrating a similar trend in their performance metrics.
The additional learnable parameters in ConSmax can lead to instability during the early stages of model training.
Moreover, the ConSmax could produce the non-unit normalization vector, which further exacerbates this instability.
That's because the non-unit normalization vector cannot effectively amplify the differences in the Q$\times$K results, causing the subsequent P$\times$V computation to fail to extract information primarily from highly relevant tokens.
Nevertheless, with sufficient training iterations, the learnable parameters in ConSmax, along with other weight matrices in LLMs, can effectively model text generation tasks.
Figure~\ref{fig:beta_gamma} illustrates the evolution of $\beta$ and $\gamma$ throughout ConSmax-based model training.
For brevity, we only present the results for one certain self-attention head in the GPT baseline model.
Similarly, $beta$ is set within the range of [0.5, 2.5], while $gamma$ is set to 100.
As the training proceeds, $\beta$ demonstrates a converging trend toward a final value.
At the same time, $\gamma$ remains relatively constant across different beta configurations.
The rest self-attention heads exhibit a similar trend.
Finally, we examine the impact of $\beta$ and $\gamma$ initialization on GPT's performance in Fig.~\ref{fig:explore}
With the same $\gamma$ value, there is a tendency for lower perplexity with smaller $\beta$ values after 10K warm-up training iterations.
Conversely, when $\beta$ values differ, the optimal $\gamma$ selection varies on a case-by-case basis.
Therefore, the combination of $\beta$ and $\gamma$ that results in the lowest perplexity is utilized to train the model until convergence.

\subsection{Hardware Performance}

\begin{figure*}[t]
\centering
\includegraphics[width=0.8\textwidth]{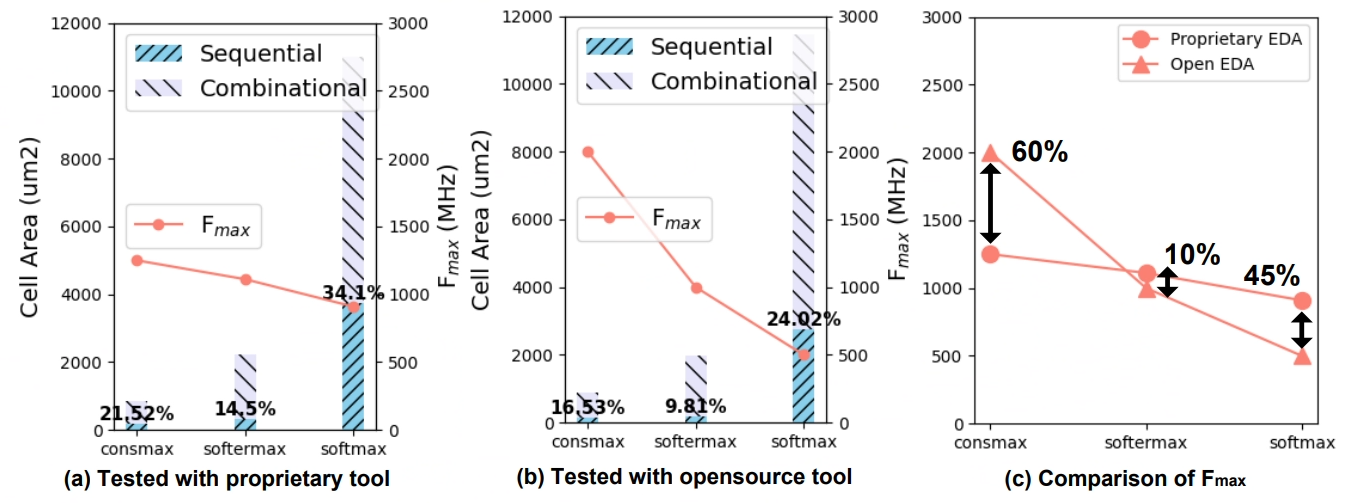}
\caption{Cell area comparison of ConSmax, Softermax and Softmax in 16nm process (a) tested in proprietary EDA tool, (b) tested in open-source EDA tool and (c) comparison of $F_{max}$ using different EDA tool for 3 designs. \label{fig:area_comparison}}
\vspace{-3mm}
\end{figure*}

\begin{figure*}[t]
\centering
\includegraphics[width=0.8\textwidth]{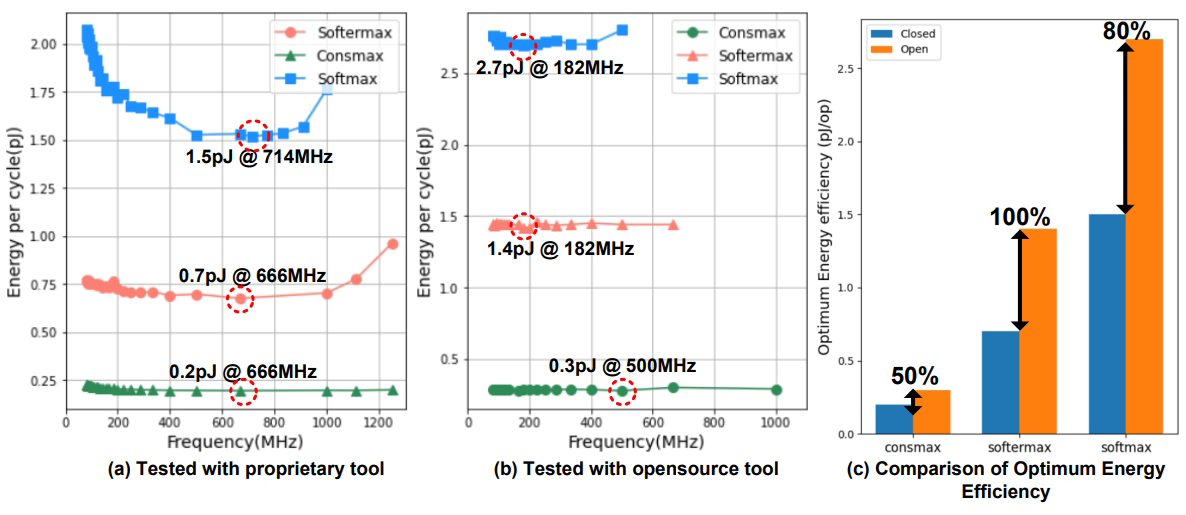}
\vspace{-2mm}
\caption{Energy efficiency comparison of ConSmax, Softermax and Softmax in 16nm process (a) tested in proprietary EDA tool, (b) tested in open-source EDA tool and (c) comparison of Energy efficiency using different EDA tool for 3 designs. \label{fig:energy_comparison}}
\vspace{-4mm}
\end{figure*}

Table~\ref{table:hardware_comparison} summarizes ConSmax hardware performance and compares it with Softmax and Softermax hardware.
For a fair comparison, the ConSmax and two baselines process a Softmax workload with a token sequence of 256.
Under 16nm FinFET technology, 1250MHz working frequency and 0.8V power supply, the ConSmax hardware only consumes 0.2mW power and 0.0008mm$\rm ^2$ area.
Therefore, the ConSmax only results in a minimal overhead when integrated into the LLM-oriented accelerators \cite{sprint, elsa, spatten}.
Compared to the Softermax counterpart, ConSmax achieves 3.35$\times$ power and 2.75$\times$ area savings.
Compared to the Softmax counterpart, ConSmax further achieves 7.5$\times$ power and 13.75$\times$ area savings.
For the open-source SkyWater's 130nm CMOS technology, ConSmax runs at a lower 667MHz working frequency and 0.8V power supply.
ConSmax achieves 3.2$\times$ power saving and 4.1$\times$ area saving compared to Softermax, and 23.2$\times$ power saving and 25.7$\times$ area saving compared to Softmax.
In summary, the ConSmax hardware presents stable performance improvement across different CMOS technologies and EDA toolchains.

Figure~\ref{fig:area_comparison} presents the area breakdown and maximum operating frequency of the three different designs.
The ConSmax demands minimal area consumption while achieving the highest operating frequency.
This is because ConSmax completely eliminates the exponential maximum and summation calculations.
As a result, the scratchpads for intermediate result storage and the costly floating-point accumulators can be minimized.
Finally, we test the energy efficiency of ConSmax on the real Softmax workload.
As depicted in Figure~\ref{fig:energy_comparison}, under 16nm CMOS technology, both ConSmax and Softermax attain optimal energy consumption at 666 MHz, whereas the Softmax baseline achieves this at 714 MHz.
ConSmax results in an energy efficiency of 0.2pJ, 3.5$\times$ and 7.5$\times$ better than Softermax and the Softmax baseline, respectively.
The open-source EDA presents a similar result.

\section{Conclusion}
\label{sec:conclusion}

This paper presents ConSmax, a software-hardware co-design for efficient Softmax acceleration.
ConSmax improves Softermax computing parallelism by avoiding data synchronization, which produces maximum score and score summation.
In addition, the bitwidth-split ConSmax hardware is lossless and scalable for calculating non-linear functions.
The experiments show that ConSmax achieves a minuscule power consumption of 0.2mW and an area of 0.0008mm$^2$ at 1250MHz working frequency and 16-nm FinFEt CMOS technology. 
Compared to state-of-the-art Softmax hardware, ConSmax results in 3.35$\times$ power and 2.75$\times$ area savings with a comparable accuracy on GPT-2 model and the WikiText103 dataset.

\bibliographystyle{unsrt}
\bibliography{references}

\end{document}